\documentclass[a4paper,useAMS,usenatbib]{mnras}
\usepackage[T1]{fontenc}
\usepackage{ae,aecompl}

\usepackage{graphicx}
\usepackage{amsmath}
\usepackage{amssymb}
\usepackage{hyperref} 
\usepackage{array} 
\usepackage{color}
\usepackage{multirow}

\newcommand{\bc}{\begin{center}}
\newcommand{\ec}{\end{center}}

\newcommand{\Msun}           {\,{\rm M}_\odot}

\title[Mergers \& Kinematically Misaligned Galaxies] {The Impact of Merging on The Origin of Kinematically Misaligned  and Counter-rotating Galaxies in MaNGA}

\author[Li S.~L. et al. ]{
Song-lin Li,$^{1,2}$
Yong Shi,$^{1,2}$\thanks{E-mail:yong@nju.edu.cn}
Dmitry Bizyaev,$^{3,4,5}$
Christopher Duckworth,$^{6,7}$
\newauthor
Ren-bin Yan,$^{8}$
Yan-mei Chen,$^{1,2}$
Long-ji Bing,$^{1,2}$
Jian-hang Chen,$^{1,2}$
Xiao-ling Yu,$^{1,2}$
\newauthor
Rogemar A. Riffel$^{9,10}$\\
$^{1}$School of Astronomy and Space Science, Nanjing University, Nanjing 210093, China.\\
$^{2}$Key Laboratory of Modern Astronomy and Astrophysics (Nanjing University), Ministry of Education, Nanjing 210093, China.\\
$^{3}$Apache Point Observatory and New Mexico State University, P.O. Box 59, Sunspot, NM, 88349-0059, USA.\\
$^{4}$Sternberg Astronomical Institute, Moscow State University, 119234, Moscow, Russia.\\
$^{5}$Special Astrophysical Observatory of the Russian AS, 369167, Nizhnij Arkhyz, Russia.\\
$^{6}$School of Physics and Astronomy, University of St Andrews, North Haugh, St Andrews, KY16 9SS, UK.\\
$^{7}$Center for Computational Astrophysics, Flatiron Institute, 162 Fifth Avenue, New York, NY 10010, USA.\\
$^{8}$Department of Physics and Astronomy, University of Kentucky, 505 Rose St., Lexington, KY 40506-0057, USA.\\
$^{9}$Departamento de F{\'i}sica, CCNE, Universidade Federal de Santa Maria, Av. Roraima, 1000, 97105-900 Santa Maria, RS, Brazil.\\
$^{10}$Laborat{\'o}rio Interinstitucional de e-Astronomia - LIneA, Rua Gal. Jos{\'e} Cristino 77, 20921-400 Rio de Janeiro, RJ, Brazil.\\
}

\date{Accepted XXX. Received YYY; in original form ZZZ}

\pubyear{2019}

\begin{document}
\label{firstpage}
\pagerange{\pageref{firstpage}--\pageref{lastpage}}
\maketitle

\begin{abstract}
Galaxy mergers and interactions are expected to play a significant role leading to offsets between gas and stellar motions in galaxies. Herein we crossmatch galaxies in MaNGA MPL-8 with the Dark Energy Spectroscopic Instrument (DESI) Legacy Surveys and identify 311 merging galaxies that have reliable measurements of the $\Delta$PA, the difference between the stellar and gas kinematic position angles to investigate the impacts of merging on gas-stellar rotation misalignments. We find that the merging fractions of misaligned galaxies (30$^\circ$ $\leqslant$ $\Delta$PA $<$150$^\circ$) are higher than that of co-rotators ($\Delta$PA $<$ 30$^\circ$) in both quiescent and star-forming galaxies. This result suggests that merging is one process to produce kinematic misalignments. The merging fraction of counter-rotators ($\Delta$PA $\leqslant$ 150$^\circ$) is lower than that of misaligned galaxies in both
quiescent and star-forming galaxies, while in the latter it is likely even lower than that of co-rotators. The orbital angular momentum transfer to the spins of stars and gas during merging and the tidal feature disappearance can lead to small merging fractions in counter-rotators. Numerous new stars that inherit angular momentum from gas after merging can further lower the merging fraction of star-forming counter-rotators.
\end{abstract}


\section{Introduction}
The merging processes have profound influences on galaxy formation and evolution, such as enhancing star formation (SF) \citep{bou11,blo17,pan19}, triggering active galactic nuclei (AGN) activity \citep{hop08,fu18}, transforming galaxy morphologies \citep{shi09,xu12,con14}, and causing chaos in the internal velocity fields of stars and gas. 

Recently, benefiting from the developments of spatially-resolved integral field spectroscopy (IFS), it is possible to measure the position angles (PAs) of the stellar and ionized gas velocity fields directly, which are inferred from the doppler shifts of the absorption lines (for stars) and the emission lines (for gas), respectively. Some galaxies exhibit offsets in the PAs ($\Delta$PA) between the stellar and gas rotation. Galaxies with 30$^\circ$ $\leqslant$ $\Delta$PA $<$150$^\circ$ are called the misaligned galaxies and those with $\Delta$PA $\geqslant$ 150$^\circ$ are counter-rotators \citep{dav11,che16,bry19}.

\cite{dav11} investigated ionized, molecular and atomic gas in 260 early type galaxies (ETGs) in ATLAS$^{3D}$ \citep{cap11}, finding a large proportion of misaligned galaxies and underlining the importance of the externally acquired gas on the gas replenishment in the ETGs. On the other hand, no striking differences in the kinematic PAs between gas and stars are found in non-interacting CALIFA galaxies \citep{san12}, which are mainly late type galaxies (LTGs) \citep{bar14}. \cite{jin16} and \cite{che16} found younger stellar populations, enhanced star formation rates and higher metallicity in central regions of misaligned galaxies in MaNGA \citep{bun15} than their outer parts, which could be caused by misaligned gas accretion and its subsequent collision with in-situ gas. \cite{bry19} (hereafter B19) made use of galaxies in SAMI \citep{bry15} covering a broad range in morphological types, stellar masses and environments, and proposed that not only gas accretion, but also the gas precession plays an important role in the apparent distributions of $\Delta$PA. In addition, gas stripping and gas disc destruction induced by AGN feedback, merging and flyby events through groups or clusters make it easier to produce gas-stellar misalignment during subsequent gas re-accretion \citep{van15,sta19,duc20b}. Meanwhile lower stellar angular momentum inherited from halo spin exerts weaker torques on gas motion, leading to a longer time for misaligned gas precessing to align with stellar motion. As a result, galaxies with lower angular momentum are more likely to exhibit such misalignment \citep{duc20a}. All of these studies share a common sense that external processes, such as gas accretion and merging, play important roles in the formation of misaligned galaxies and counter-rotators.
 
It is expected that the misaligned and counter-rotating gas origins from external processes such as merging as seen in simulations \citep{van15}. The observational evidence for this has emerged from \cite{bar15}. Based on 66 interacting galaxies that have well defined $\Delta$PA (out of 103 total interacting galaxies), they found interacting galaxies have larger kinematic misalignments than non-interacting galaxies. With a sample of five times larger than theirs, we will further investigate the effect of merging in producing misaligned galaxies and counter-rotators, and especially we will discuss the effect in different groups of galaxy types including star-forming and quiescent galaxies. 

Mapping Nearby Galaxies at Apache Point Observatory (MaNGA) \citep{bun15} will finish observations of 10000 nearby galaxies designed to have a flat stellar mass distribution from $\rm 10^9 \Msun$ to $\rm 10^{11} \Msun$ with IFS data until 2020 \citep{wak17}. The latest internal data release MPL-8 contains 6505 unique galaxies. Meanwhile, there are plenty of methods to identify mergers, such as tidal features \citep{mar10,wen14,hoo18,mor18}, morphological asymmetries \citep{rei08,shi09,lot10,con14} or the galaxy pairs \citep{ell08,li08,tor12}. All of them can be extracted from the optical images. By combining MaNGA with imaging surveys, and looking for merger vestiges, we are able to directly examine the importance of the role that merger plays in kinematically misaligned galaxies. 

The paper is organized as follows. In Section 2, we briefly introduce the MaNGA project, Legacy Surveys, sample selection and methodology. The observational results are shown in Section 3. We discuss the possible explanation for the $\Delta$PA distribution of our sample in Section 4. A summary and conclusion are listed in section 5.

\section{Data and Methodology}
\subsection{MaNGA}
MaNGA, which started in 2014 July using the 2.5 meters telescope at Apache Point Observatory (APO) \citep{gun06}, is one of three major programs in Sloan Digital Sky Survey IV (SDSS-IV, \citep{bla17}). This program aims to acquire integral field unit (IFU) spectra for 10000 nearby galaxies with redshift ranging from 0.01 to 0.15 \citep{wak17}. Each target is observed to ensure the $\rm 5\sigma$ depth to reach 23 mag $\rm arcsec^{-2}$ \citep{law15}. The coverage of each galaxy is expected to be 1.5 effective radii ($R_e$) for primary sample and 2.5 $R_e$ for secondary sample with typical angular resolution $\thicksim$ 2.5 arcsec, corresponding to 1$\thicksim$2 kpc in such redshift range \citep{yan16b}. Thus the bundles are designed to contain 19 to 127 fibers to satisfy these coverages of galaxies with different angular size \citep{dro15}, with the mini-bundles with 7 fibers for flux calibration \citep{yan16a}. The dual-channel BOSS spectrographs cover a wavelength range of 3600$\thicksim$10300 \AA $~$with spectral resolution R$\thicksim$2000 \citep{sme13}. The raw data is reduced through data reduction pipeline (DRP) \citep{law16} and analyzed through Data Analysis Pipeline (DAP) \citep{wes19,bel19}. In the latest internal data release MPL-8, there are 6505 unique galaxies with 3D data cubes and 2D data maps, which is the largest IFU survey sample to date.

\subsection{Legacy surveys}
The Dark Energy Spectroscopic Instrument (DESI) Legacy Imaging Surveys (hereafter Legacy Surveys) \citep{dey19} aim to provide targets for the DESI survey. The Legacy Surveys are a combination of three public projects covering about 14000 $\rm deg^2$ of the sky visible from the northern hemisphere: the Beijing-Arizona Sky Survey (BASS) \citep{zou17} observed by the 90Prime camera \citep{wil04} on the Bok 2.3-meter telescope on Kitt Peak; the Mayall $z$-band Legacy Survey (MzLS) \citep{sil16} observed by the Mosaic3 camera \citep{dey16} on the 4-meter Mayall telescope at Kitt Peak and the Dark Energy Camera Legacy Survey (DECaLS) \citep{blu16} by the Dark Energy Camera \citep{fla15} on the 4-meter Blanco telescope at the Cerro Tololo Inter-American Observatory. BASS covers about 5400 $\rm deg^2$ in the north Galactic cap (dec$>$32$^\circ$) providing $g$- and $r$-band images and MzLS complements the same region as BASS with $z$-band observations. In tandem, DECaLS surveys an equatorial area (dec$<$32$^\circ$) of about 9000 $\rm deg^2$ with $g$-, $r$- and $z$-band images. The surface brightness (SB) limits of BASS are $g$=27.10 and $r$=26.76 mag arcsec$\rm^{-2}$ (5$\sigma$ in 10$\times$10 arcsec boxes). The SB limits of DECaLS are $g$=27.77 and $r$=27.44 mag arcsec$\rm^{-2}$. They are about 1 mag deeper than those of SDSS. These deeper images will facilitate us to search for debris features of merging and strongly interacting galaxies.
 
\subsection{Methods}
{\bf Sample selection:} We crossmatch the galaxies in MaNGA MPL-8 with legacy surveys\footnote{http://legacysurvey.org} DR7 (DECaLS) and DR6 (BASS DR2\footnote{http://batc.bao.ac.cn/BASS/doku.php?id=home} \citep{zou18}). Since the BASS data are independently processed by the BASS team and the image quality is better than sections provided from the Legacy Surveys data release, we use BASS images rather than DECaLS ones for those available in both surveys. Bad images including those with bad pixels across the galaxy, those with distortion during mosaic and those at the edge of an image segment, are eliminated from the sample. Finally we get 6217 galaxies (hereafter crossmatched MaNGA sample) with 3729 from BASS and 2488 from DECaLS. All of them have both $g$- and $r$-band images. For each galaxy we convolve one band that has smaller seeing with a Gaussian function to match the PSF in the other band and stack them to enhance S/N. Then we visually check all those stacked images in DS9 to visually identify faint features.

We finally find 538 merging/interacting galaxies and roughly divide them into four groups, as illustrated in Fig. 1. The first group contains isolated galaxies with tidal streams (219, top left). The second group is composed of distorted galaxies that have companions (184, top right). This type of galaxies belongs to galaxy pairs but pairs without strong interacting features, i.e. bridges, are not included in this group. The third group consists of galaxies with shells (36, bottom left). Shells are very faint structures, so we only identify 36 galaxies with shells. Here we include two types of shells as discussed in \cite{ebr13}. Type I is cone like and interleaved in radius. Type II is randomly distributed arcs. The unique feature of shells is ripple like, so that galaxies with extended symmetric smooth halos are not in our sample. The fourth group contains galaxies with extended asymmetric halos (99, bottom right). Thus except for the second group, all the merging galaxies are remnants and most of them may have experienced major mergers. By adopting the above four groups of merging/interacting features, we can better train our eyes to have fair classifications. We will not discuss the difference among different groups in this work, and use all of them as a whole.

\begin{figure*}
  \resizebox{15cm}{!}{\includegraphics{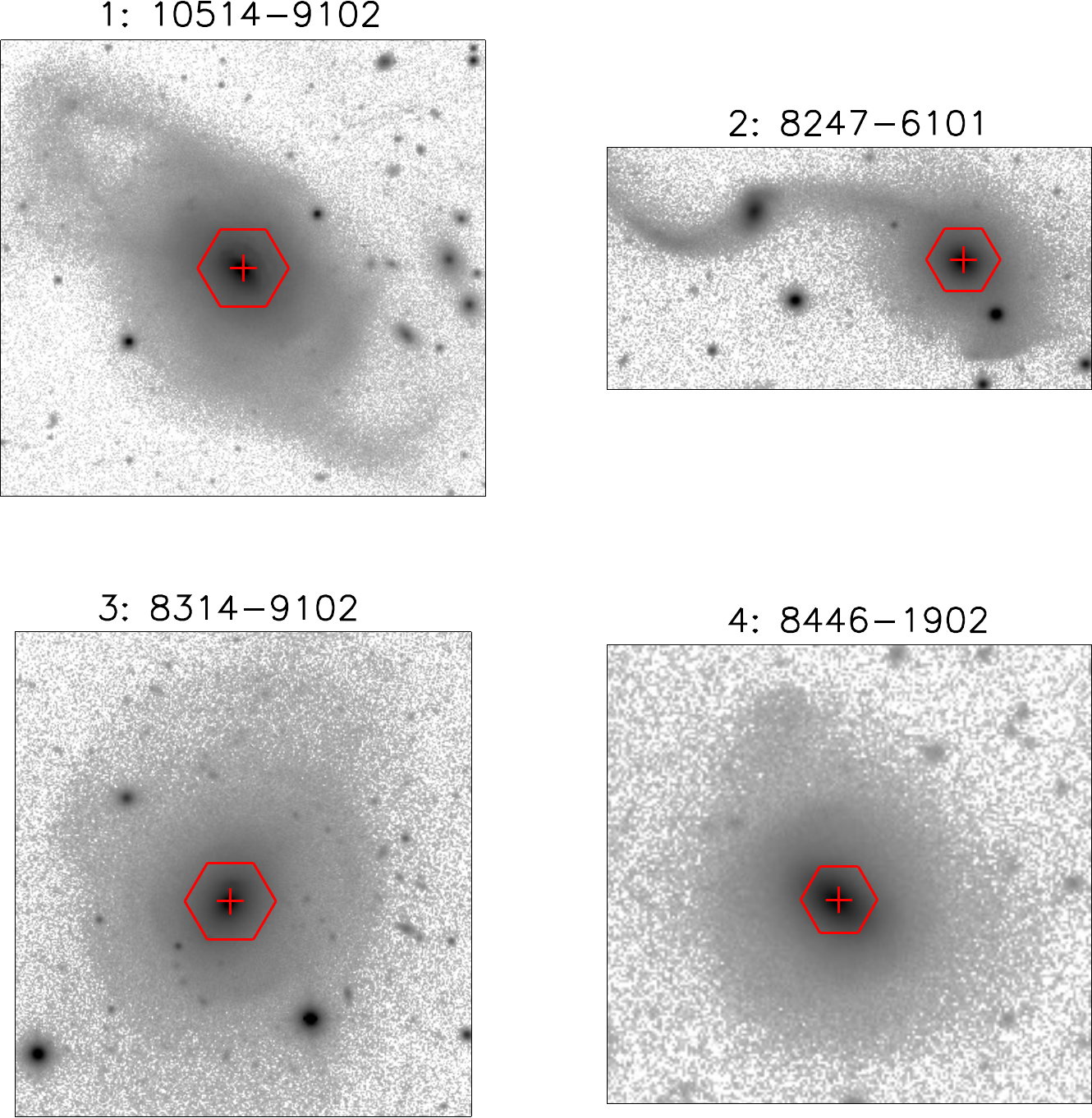}}
  \caption{Four examples from different groups of our sample. Those are: (1) isolated galaxies with tidal features (top left); (2) distorted galaxies with companions (top right); (3) galaxies with shells (bottom left) and (4) galaxies with extended asymmetric halo (bottom right). The plate-ifu IDs in MaNGA are showed on the top of each panel. Red plus symbol marks the center of each galaxy. The MaNGA IFU footprint is also overlaid in red. }
\end{figure*}

{\bf Stellar mass and star formation rate:} The total stellar mass and SFR of the crossmatched MaNGA sample are obtained from two catalogs - GALEX-SDSS-WISE Legacy Catalog (GSWLC) from \cite{sal16} and MPA-JHU DR7 catalog\footnote{https://wwwmpa.mpa-garching.mpg.de/SDSS/DR7/}. GSWLC derived physical properties from the UV/optical SED fitting, which is robust but only contains 5109 out of 6217 galaxies. In addition, we crossmatch the rest of them with MPA-JHU DR7 to add 840 more objects. The stellar mass and SFRs for star forming galaxies in this catalog are calculated following \cite{kau03} and \cite{bri04}, respectively. For non-SF galaxies, the SFRs are estimated from D4000. The two catalogs agree with each other quite well in stellar mass and SFRs for SF galaxies, but the MPA catalog tends to overestimate the SFRs slightly for non-SF galaxies. Finally, there are 268 crossmatched MaNGA sample without stellar mass and SFR measurements. We don't use them for the following investigations. Fig. 2 shows the SFR-M$_*$ diagram of our merging/interacting sample in MaNGA.

\begin{figure*}
  \resizebox{13cm}{!}{\includegraphics{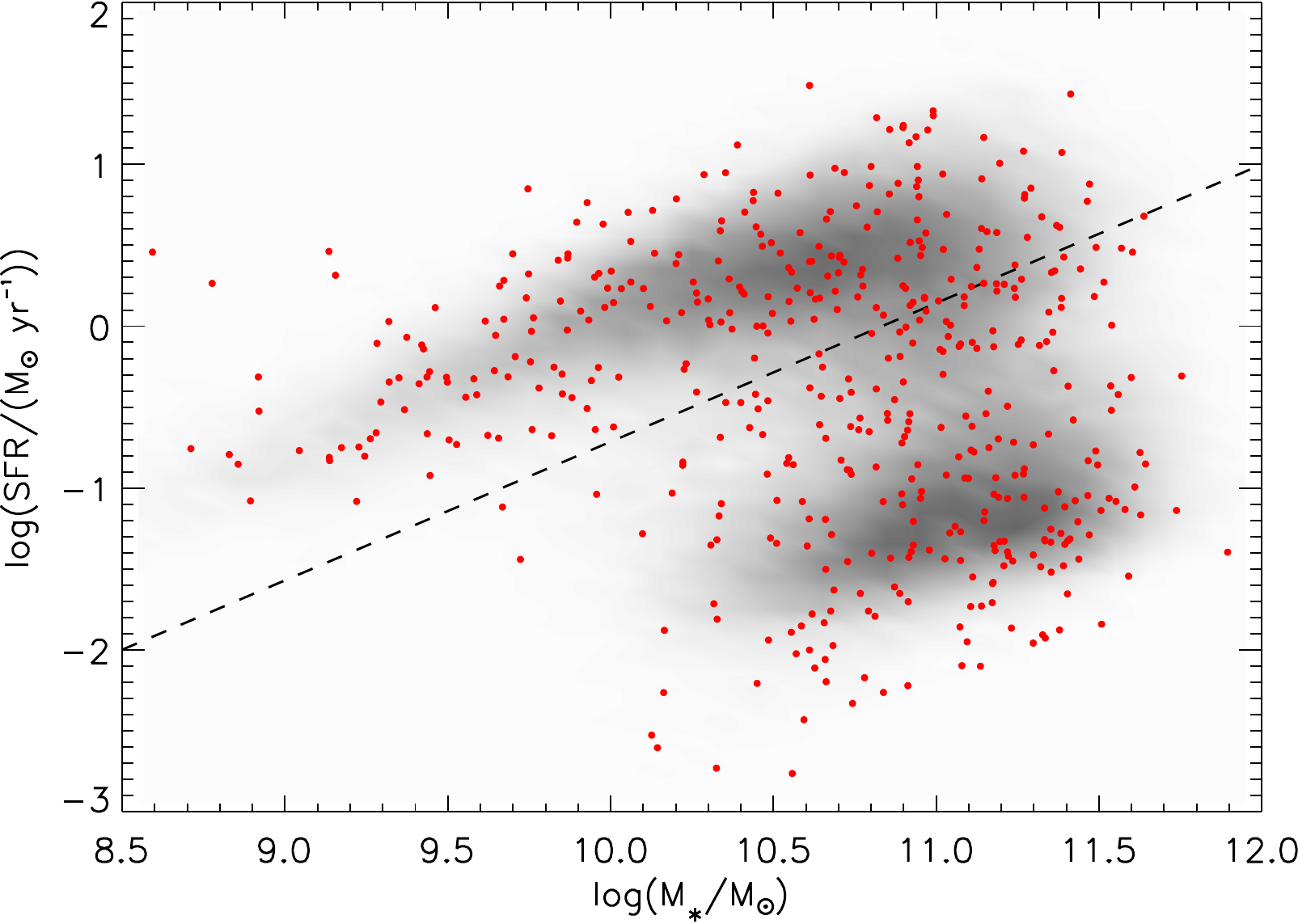}}
  \caption{{\bf (a):} The distribution of our sample in SFR-M$_*$ diagram. Red dots are 485 of merging galaxies having both SFR and stellar mass measurements from literatures. Shadows in the background show the distribution of 100 thousand galaxies from GSWLC catalog. The dashed line is the 1$\rm \sigma$ lower boundary of star forming main sequence from \citet{che16}.}
\end{figure*}

{\bf PAs of stellar and gas rotations:} The velocity maps of both stellar and ionized gas are extracted from the output of data analysis pipeline (DAP) in MaNGA \citep{wes19,bel19}. We are using VOR10 velocity maps from DAP throughout this paper. The pipeline first bins adjacent spaxels and stacks the spectra in those spaxels using Voronoi binning procedure \citep{cap03} to ensure the pseudo-$r$-band S/N bigger than 10. Then the continuum of each bin is fitted using penalized pixel-fitting routing (pPXF) \citep{cap04,cap17} and hierarchically clustered MILES templates (\texttt{MILES-HC}) \citep[MILES stellar library: ][]{san06} to determine the stellar kinematics. After performing stellar-continuum fit, the pipeline fixes the stellar kinematics and fits the emission lines by adopting Gaussian emission-line models and continuum simultaneously, providing best fit continuum models and fluxes and equivalent widths of emission lines, as well as velocities and velocity dispersions. 
 
The PAs of stellar and gas rotations are measured using \texttt{FIT\_KINEMATIC\_PA} code\footnote{https://www-astro.physics.ox.ac.uk/~mxc/software/}, which is based on algorithms proposed in Appendix C of \cite{kra06}. To obtain robust PAs of both stellar and gas, we apply following criteria: (1) only spaxels with \texttt{mask}=0 in DAP output are used, which means there are no significant issues of the data quality and the fit is reasonable; (2) for stellar rotation, we eliminate spaxels with velocity uncertainties $>$ 30 km/s. As for gas, we only include spaxels after binning with fluxes S/N of either $\mathrm{H\alpha}$ or $\mathrm{O[III] \lambda 5007}$ greater than 5; (3) the number of spaxels satisfying the above second criterion needs to be greater than 1/3 of spaxels with \texttt{mask}=0. We then visually check each velocity map with successful PA fitting to ensure that the measurements are reasonable.

The advantage of \texttt{FIT\_KINEMATIC\_PA} code is that it is not sensitive to the local distortion of the velocity fields. To ensure that our PA measurements are robust, we visually check every velocity map. We find that two cases could make the fitted PA deviate from the real rotation direction. The first case is the spuriously local high velocities that might be attributed to background targets. This high-velocity spaxels contribute to the $\chi^2$ significantly and thus tend to make the fitted PA along their direction. We mask corresponding region and refit the velocity maps. The second case is that the procedure could not fit edge-on galaxies well as also found in \cite{duc20a}. Therefore, we eliminate those edge-on galaxies with obvious deviations of PA measurements from the real PA of either stellar or gas velocities. 

At the end 3681 galaxies out of 6217 have both reliable stellar and gas PAs, while most of the rest either don't have enough gas to produce measurable emission lines or don't have coherent gas or stellar velocity fields and thus don't have an overall PA. Among merging/interacting galaxies, 311 out of 538 have PA measurements available. The PAs are in the range from 0$^\circ$ to 360$^\circ$, so we force the PA offsets between gas and stellar from 0$^\circ$ to 180$^\circ$.  

\section{Results}	 
Following the classification in \cite{che16}, galaxies with $\Delta$PA $<$ 30$^\circ$ are co-rotators or aligned galaxies; those with $\Delta$PA $\geqslant$ 150$^\circ$ are counter-rotators and the rest (30$^\circ$ $\leqslant$ $\Delta$PA $<$150$^\circ$) are misaligned galaxies (Here the definition is a little bit different with \cite{che16}, where counter-rotators are also misaligned galaxies). Fig. 3 shows examples of the velocity maps and measured PA of one co-rotator (top row), one misaligned galaxy (middle row), and one counter-rotator (bottom row). We further divide our sample into star-forming and quiescent galaxies according to the dashed line in Fig. 2. Table 1 lists the statistics on the numbers of galaxies in each $\Delta$PA group.

\begin{table*}
\begin{center}
\caption{The number of galaxies in three kinematic groups. Each group is further separated into star-forming galaxies and quiescent galaxies. Fig. 3 shows examples of these three kinematic groups.}
\begin{tabular}{lc|ccc}
\hline 
\multicolumn{2}{c|}{Groups}& Co-rotators  & Misaligned galaxies  & Counter-rotators \\ 
&& (0$^\circ$-30$^\circ$)& (30$^\circ$-150$^\circ$) & (150$^\circ$-180$^\circ$) \\\hline
&Star-forming & 179 & 13 & 2 \\
Merging/interacting&Quiescent& 80 & 29 & 8 \\
&Total & 259 & 42 & 10 \\\hline
&Star-forming & 2420 & 101 & 39 \\
The whole&Quiescent & 915 & 136 & 70 \\
&Total & 3335 & 237 & 109 \\\hline
\end{tabular}
\end{center}                                                                                                       
\end{table*}

\begin{figure*}
 \resizebox{16cm}{!}{\includegraphics{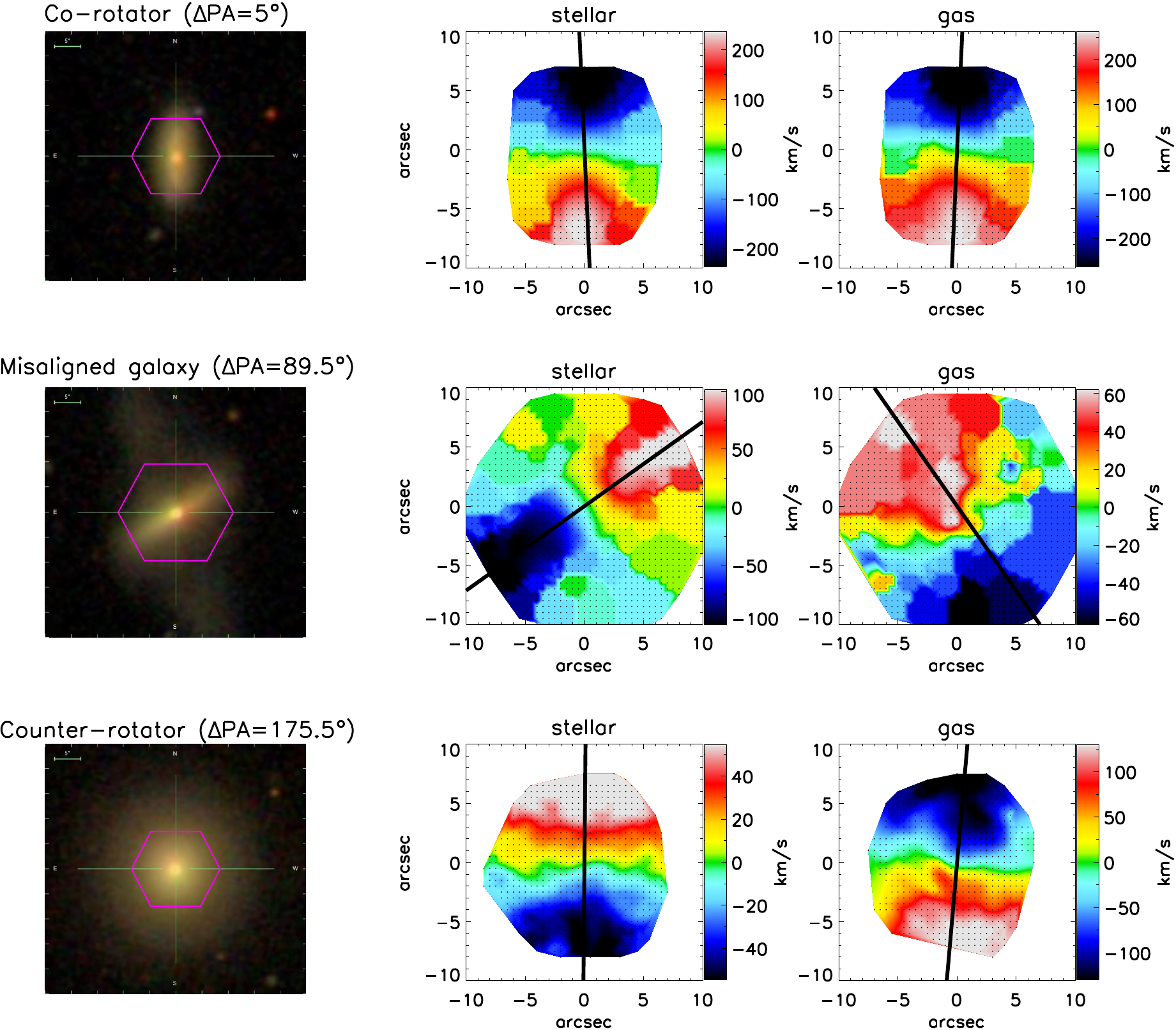}}
 \caption{Three examples of a co-rotator (first row), a misaligned galaxy (second row) and a counter-rotator (third row). The first column shows SDSS images with MaNGA field of view overlapped in magenta. The second and third columns show stellar velocity fields and gas velocity fields, respectively. Major axes are illustrated as black lines.}
\end{figure*}

\subsection{The $\Delta$PA distributions}
Fig. 4 shows the $\Delta$PA distributions of star-forming galaxies (blue bars) and quiescent galaxies (red bars). In both groups, while co-rotators are dominant, 18\% of quiescent galaxies have $\Delta$PA$\geqslant$ 30$^\circ$, but only 5\% of star-forming galaxies do. B19 find that only 5\% of star-forming galaxies that are morphologically identified have $\Delta$PA greater than 30$^\circ$ in SAMI sample and \cite{duc20a} identify misaligned galaxies and counter-rotators to be 5.4\% of morphologically-identified star-forming galaxies in MaNGA. These works all find that gas-rich galaxies are hard to produce stars and gas kinematic misalignment and are in line with our results. We thus draw a same conclusion with these works but with six times larger sample size than B19. According to B19, in star-forming galaxies, the interaction between existing co-rotating gas and incoming gas rapidly disrupts the latter and produces a co-rotator even if the incoming gas is misaligned with stellar rotation. Moreover, simulations of halo accretion \citep{dan15} show that the accreted gas in star-forming galaxies is tilted rapidly toward to stellar rotation outside 0.1 virial radii, such that the external gas in the region covered by MaNGA is more likely to already align with stellar rotation. Both co-rotating settling and preferentially aligned accretion result in very few misaligned galaxies and counter-rotators in star-forming galaxies.

\begin{figure*}
 \resizebox{13cm}{!}{\includegraphics{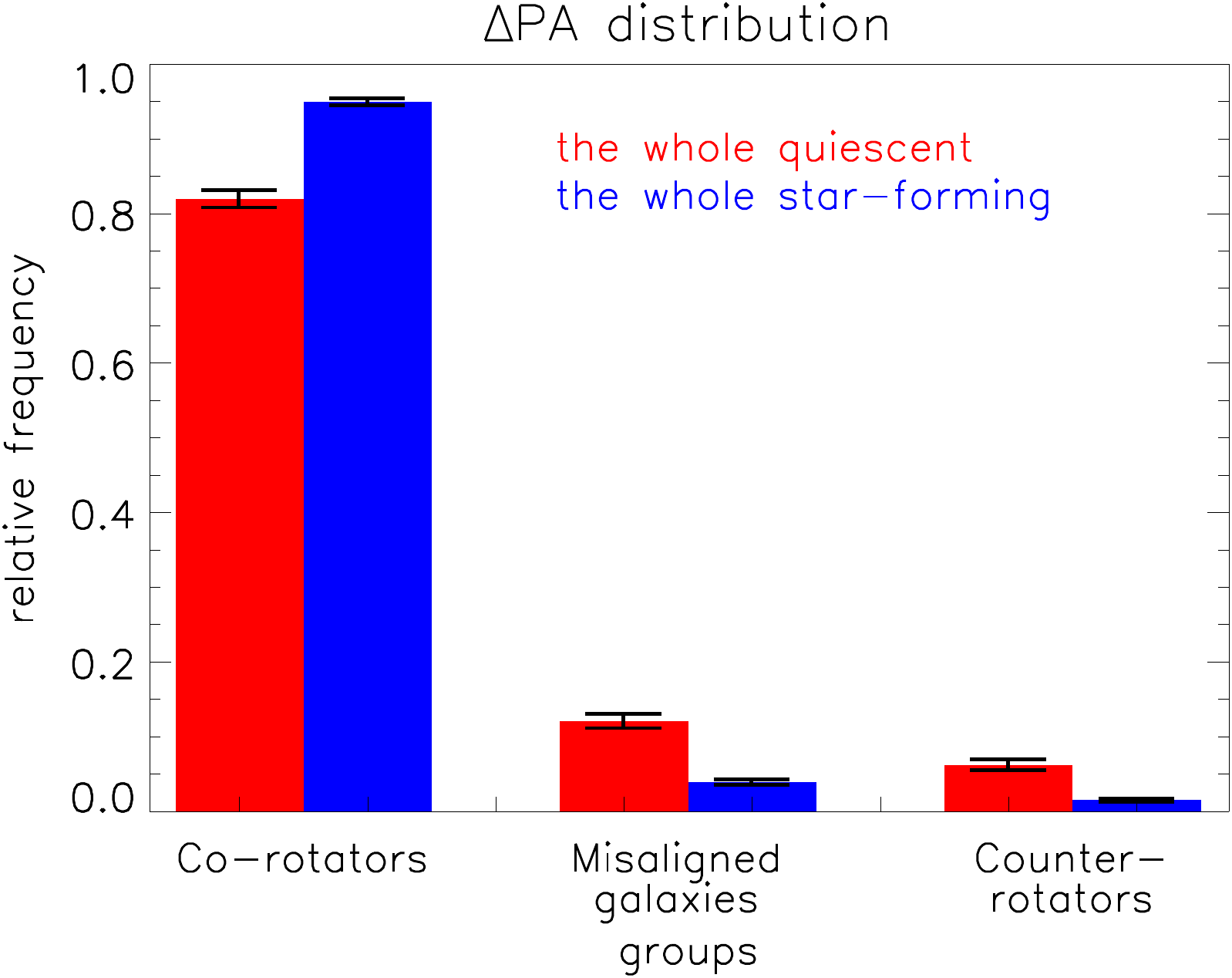}}
 \caption{The $\Delta$PA distributions of the whole quiescent galaxies (red bars) and the whole star-forming galaxies (blue bars). Co-rotators are those with $\Delta$PA$<$30$^\circ$; misaligned galaxies are those with 30$^\circ$$\leqslant$$\Delta$PA$<$150$^\circ$; and counter-rotators are those with $\Delta$PA $\geqslant$150$^\circ$. The error bars illustrate the square roots of the variances of the $\beta$ distributions.}
\end{figure*}

The relative number ratio of misaligned galaxies to counter-rotating galaxies is 2:1 and 2.5:1 in quiescent and star-forming galaxies, respectively. This is smaller than the simple expectation from the $\Delta$PA coverage of these two groups, which should be 120$^\circ$:30$^\circ$=4:1. This means that we find more counter-rotators than predicted. B19 also find this counter-rotator excess and they use gas precession scenario to explain it. Due to gravitational dynamical settling, misaligned gas will precess to be stabilized in either counter-rotating, if the initial $\Delta$PA is bigger than 90$^\circ$, or co-rotating, if the initial $\Delta$PA is smaller than 90$^\circ$, with stellar motion \citep{ste16,dav16}. The precessing timescale depends on both ellipticity of the existing stellar disc and the gas inclination to the galaxy. For example, it will take over Hubble time for one near polar gas disc in round ellipticals (e.g.: type E3 with b/a=0.7) precessing and aligning with stellar motion \citep{sch83}. Gas precession alters some misaligned galaxies to counter-rotators, thus resulting in more counter-rotators.

\subsection{The merging fractions}
Fig. 5 shows the merging fractions of the whole sample (green bars), the quiescent galaxies (red bars), and the star-forming galaxies (blue bars). It is evident that misaligned galaxies have a higher merging fraction than co-rotators. This excess is seen in both quiescent (the difference between misaligned galaxies and co-rotators is 12\%$\pm$3.6\%) and star-forming galaxies (the difference is 6\%$\pm$3.3\%), and of course the whole sample (the difference is 10\%$\pm$2.5\%). We have also run FIT\_KINEMATIC\_PA on unbinned stellar and gas velocity maps. Then we use different stellar and gas map combinations to calculate the $\Delta$PA and this excess still exists as the merging fractions of co-rotators and misaligned galaxies are 7.7\%$\pm$0.5\% and 18.2\%$\pm$2.6\% for unbinned stellar and binned gas maps, 7.7\%$\pm$0.5\% and 19.6\%$\pm$2.9\% for binned stellar and unbinned gas maps, and 7.8\%$\pm$0.5\% and 19.1\%$\pm$3.0\% for unbinned stellar and unbinned gas maps. \cite{bar15} also find that interacting galaxies have higher mean $\Delta$PA compared with non-interacting galaxies. Our results show that this elevated merging fractions of misaligned galaxies are independent of the SFRs. This supports that merging is responsible for the origin of misaligned galaxies in both star-forming galaxies and quiescent galaxies. MaNGA FoV only covers to 1.5 Re or 2.5 Re of each galaxy. According to B19, the precession timescale depends on the radial distance to the galactic center, with gas in the lower radius precessing faster and aligning with stellar rotation. Therefore, for warped disc galaxies \citep{deb99,rad14,spa09,van15}, if an IFU with FoV larger than MaNGA is adopted, some co-rotators will be classified as misaligned galaxies, enhancing the merging fraction of misaligned galaxies. Our results remain unchanged.

\begin{figure*}
  \resizebox{13cm}{!}{\includegraphics{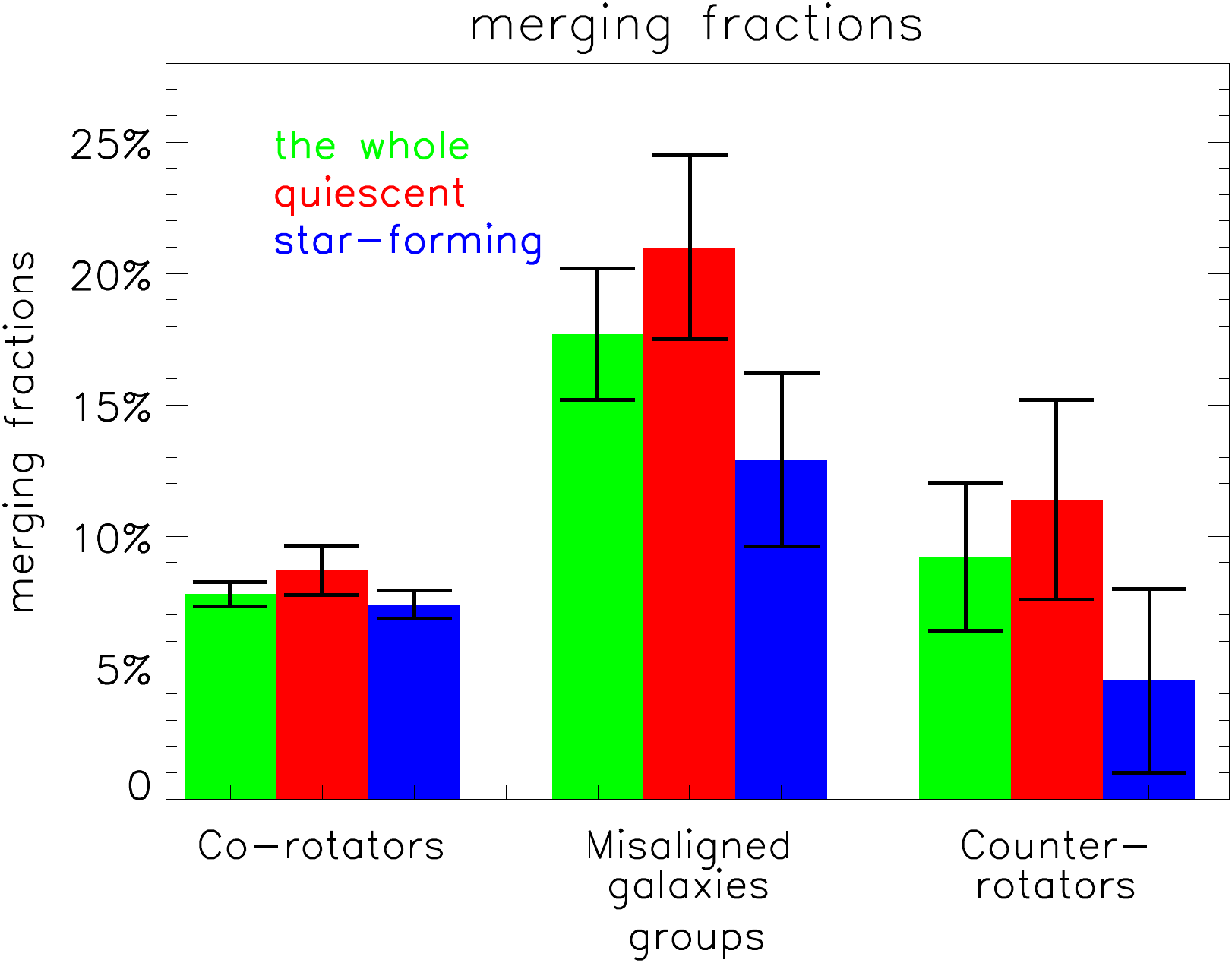}}
  \caption{The fractions of merging/interacting galaxies in each kinematic group. Green bars represent the sum of quiescent and star-forming galaxies, Red bars represent quiescent galaxies, and blue bars represent star-forming galaxies. Co-rotators are those with $\Delta$PA$<$30$^\circ$; misaligned galaxies are those with 30$^\circ$$\leqslant$$\Delta$PA$<$150$^\circ$; and counter-rotators are those with $\Delta$PA $\geqslant$150$^\circ$. The error bars illustrate the square roots of the variances of the $\beta$ distributions.}
\end{figure*}  

The merging fraction of counter-rotators is smaller than that of misaligned galaxies too, with the difference of 9\%$\pm$3.8\% in the whole sample and 10\%$\pm$5.2\% and 9\%$\pm$4.8\% in quiescent and star-forming galaxies, respectively. In star-forming galaxies, the merging fraction of counter-rotators is even smaller than that of co-rotators in spite of statistical insignificance. This indicates that merging is hard to produce star-forming counter-rotators.

\subsection{The mass effects}
Next, we investigate whether misaligned galaxies have a dependency on stellar mass and whether this dependency would change our main result that misaligned galaxies generally have higher merging fractions than co-rotators.

Fig. 6 shows the stellar mass distributions of those with $\Delta$PA$<$ 30$^\circ$ and those with $\Delta$PA$\geqslant$ 30$^\circ$ in the whole MaNGA sample. We find galaxies with $\Delta$PA$\geqslant$ 30$^\circ$ tend to have smaller stellar mass than those with $\Delta$PA$<$ 30$^\circ$ in both star-forming and quiescent sample, with the KS test p-value=8.62$\rm\times 10^{-8}$ for star-forming galaxies and 4.02$\rm\times 10^{-8}$ for quiescent galaxies. \cite{duc20a} find the same trend in star-forming galaxies but an inverted trend in quiescent galaxies. The reason of this difference is that we use a different criterion. Their quiescent galaxies are chosen by morphologies. We find that their stellar mass distribution of ETGs is bimodal, with the low mass peak around 10$^{9.1} \Msun$. This peak can not be found in Fig. 6(b). We crossmatch their sample with ours and reveal that 90\% of their ETGs with stellar mass lower than 10$^{9.8} \Msun$ are actually star-forming galaxies in our sample. The stellar mass of co-rotators around this peak is smaller than that of misaligned galaxies, which leads to the overall higher stellar mass of misaligned galaxies. If we only inspect the distribution with ETGs stellar mass higher than 10$^{9.8} \Msun$ (Fig. 11 in \cite{duc20a}), misaligned galaxies are truly less massive than co-rotators, in line with our results.

We then draw from the co-rotators to create a subsample with a consistent mass distribution as those with $\Delta$PA$\geqslant$ 30$^\circ$. We find that the merging/interacting fractions are 5.8\% for star-forming co-rotators and 7.2\% for quiescent co-rotators, showing more difference against other two groups. Hence we confirm that the kinematic misalignment enhancement in the merging/interacting sample is really attributed to mergers but not stellar mass.   

\begin{figure*}
  \resizebox{16cm}{!}{\includegraphics{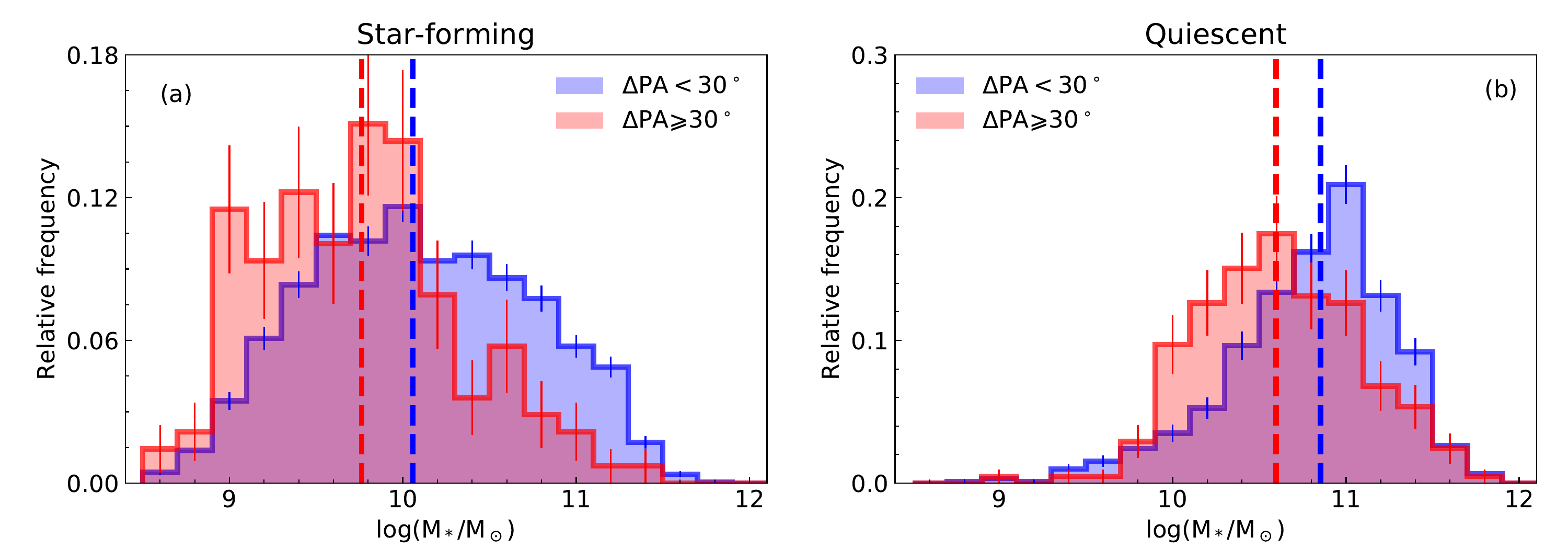}}
  \caption{The stellar mass distributions of the whole star-forming galaxies (a) and quiescent galaxies (b) in MaNGA sample. The red lines illustrate those with $\Delta$PA$\geqslant$30$^\circ$ and the blue lines illustrate those with $\Delta$PA$<$30$^\circ$. The vertical dashed lines are the median value of each distribution. Error bars show the square roots of the variances of the $\beta$ distributions in each bin.}
\end{figure*}

\section{Discussion}
In this work, our main finding is that misaligned galaxies have higher merging/interacting fraction than co-rotators. This phenomenon exists in both quiescent galaxies and star-forming galaxies and provides us with a direct observational evidence that merging is one way to produce kinematically misaligned galaxies. We do caution that our heavy Voronoi binning and symmetrisation of the velocity fields with kinematic perturbations can potentially cause large errors in the estimations of PA at least for some cases, although we have done visually check.

As shown in Fig. 5, and listed in Table 1, the merging/interacting fraction of the whole sample is about 8\%, whereas the merging/interacting fraction of misaligned galaxies is about 18\%. This fraction will be changed if the depth of images is deeper or the methods to identify merging are different. In previous works, different merging tracers reveal that merging fraction in local universe is no more than 20\% \citep{lef00,cas05,lot08,shi09}. If we adopt this upper limit and keep the ratio of merging fractions between misaligned galaxies and the whole sample, then the maximum merging fraction in misaligned galaxies is about 45\%, whereas other misaligned gas can originate from gas accretion from cosmic web or surrounding gas-rich satellites, which can not be identified by photometry in optical bands.

In the next, we will discuss the results of quiescent galaxies and star-forming galaxies separately.

\subsection{Quiescent galaxies}
As shown in Table 1, the ratio of misaligned galaxies to counter-rotators in quiescent merging/interacting sample is about 4:1, greater than that ratio in the whole quiescent sample (2:1 as discussed above). We think this is because during merging galaxy pairs transfer their orbital angular momentum into spin of remnant halo, and also of stars and gas \citep{dim09,cod12,ste17,lag18}. In semi-analytic model, the spin direction of merger remnant is set as the orbital axis direction of merging galaxies in the last resolved moment \citep{ste16}. Orbital angular momentum transfer during the merger reduces the overall $\Delta$PA so that some progenitors with gas and star initially counter-rotating end up as misaligned remnants. This effect also reduces the merging fraction in counter-rotators as seen in Fig. 5. In addition,  counter-rotating is a stable phase \citep{bry19,ste16}. Tidal features in merging/interacting counter-rotators will disappear or immerse into noise in this stable configuration and fail to be identified from optical images. It is likely that both the orbital angular momentum transfer during the merger and the tidal feature disappearance lead to a smaller merging fraction of counter-rotators than misaligned galaxies.

\subsection{Star-forming galaxies}
Fig. 5 shows a hint that the merging fraction of star-forming counter-rotators is low, even smaller than that of star-forming co-rotators. This suggests that merging plays a minor role in producing star-forming counter-rotators. B19 also indicated that star-forming counter-rotators are unlikely a production of merging based on the marginal dependency on local environment density. \cite{alg14} alternately proposed that a specific 'V-shaped' gas accretion from cosmic filaments could produce star-forming counter-rotators, with no merging process.
 
The minor effects of merging on producing star-forming counter-rotators could be also attributed to the disappearance of tidal features and the orbital angular momentum transfer as explained for quiescent counter-rotators. Table 1 shows that the ratio between star-forming misaligned galaxies and star-forming counter-rotators in merging/interacting sample is 7:1, greater than that ratio in quiescent merging sample (4:1). This requires more efficient angular momentum transfer. In addition, \cite{lag18} use EAGLE simulation and find out that wet mergers spin up galaxies. The increased stellar angular momentum originates from new formed stars out of gas disc formed after merging. These new born stars inherit angular momentum of the gas, and thus form a co-rotator. Stellar angular momentum inheritance also make it hard to form a star-forming counter-rotator after merging.

\section{Summary}
In 6217 MaNGA galaxies crossmatched with Legacy Surveys, we find 538 galaxies with merging features or strong interaction with companions. They can be roughly separated into four groups - isolated galaxies with tidal streams (219); distorted galaxies with companions (184); galaxies with shells (36) and galaxies with extended asymmetric halo (99). Among these galaxies, 3681 out of 6127 galaxies have both reasonable stellar and gas PA measurements, and 311 out of 538 merging/interacting sample does. Then we separate the whole sample into quiescent galaxies and star-forming galaxies.

We firstly investigate the gas-stellar rotation misalignments distribution in quiescent galaxies and star-forming galaxies. We find that 18\% quiescent galaxies have $\Delta$PA $\geqslant 30^\circ$ and only 5\% star-forming galaxies do, which shows that star-forming galaxies are hard to produce kinematic misalignments. Our results are consistent with former works and we corroborate them with a sample size six times larger than former works.

Next, we investigate the merging fraction in three $\Delta$PA groups and find that the merging fractions of misaligned galaxies are higher than that of co-rotators in both quiescent and star-forming galaxies. This result shows a direct evidence that merging is one process to produce stellar-gas misalignments. In addition, the merging fraction of counter-rotators is smaller than misaligned galaxies. Both the orbital angular momentum transfer during the merger and the tidal feature disappearance can lead to this smaller merging fraction. For star-forming galaxies, we find a hint that the merging fraction of counter-rotators could be even smaller than that of co-rotators. Apart from two possible explanation mentioned above, we think new stars inheriting angular momentum from gas also make it hard to form a star-forming counter-rotator after merging.
 
\section{Acknowledgements}
We thank the referee for a detailed report that helped significantly in improving the presentation of our work. S.L. and Y.S. acknowledge the support from the National Key R\&D Program of China (No. 2018YFA0404502, No. 2017YFA0402704, ), the National Natural Science Foundation of China (NSFC grants 11825302, 11733002 and 11773013). D.B. is supported by grant RScF 19-12-00145. C.D. acknowledges support from the Science and Technology Funding Council (grant number ST/N504427/1). The Flatiron Institute is supported by the Simons Foundation. Y.C. acknowledges support from the National Key R\&D Program of China (No. 2017YFA0402700), the National Natural Science Foundation of China (NSFC grants 11573013, 11733002, 11922302). R.A.R thanks partial financial support from Conselho Nacional de Desenvolvimento Cient\'ifico e Tecnol\'ogico (202582/2018-3 and 302280/2019-7) and Funda\c c\~ao de Amparo \`a pesquisa do Estado do Rio Grande do Sul (17/2551-0001144-9 and 16/2551-0000251-7). Funding for the  Sloan Digital Sky Survey IV has  been provided by the Alfred P.  Sloan Foundation, the  U.S. Department of Energy  Office of Science,  and the  Participating Institutions.  SDSS- IV  acknowledges support and  resources from the Center  for High-Performance Computing at the University of Utah. The SDSS web site is www.sdss.org. SDSS-IV is  managed by the  Astrophysical Research Consortium  for the Participating  Institutions of  the SDSS  Collaboration including  the Brazilian Participation  Group, the Carnegie Institution  for Science, Carnegie  Mellon  University,  the Chilean  Participation  Group,  the French   Participation    Group,   Harvard-Smithsonian    Center   for Astrophysics,  Instituto de  Astrof\'{i}sica  de  Canarias, The  Johns Hopkins University, Kavli Institute for the Physics and Mathematics of the Universe (IPMU) / University  of Tokyo, Lawrence Berkeley National Laboratory,  Leibniz  Institut   f\"{u}r  Astrophysik  Potsdam  (AIP), Max-Planck-Institut    f\"{u}r     Astronomie    (MPIA    Heidelberg), Max-Planck-Institut     f\"{u}r     Astrophysik    (MPA     Garching), Max-Planck-Institut f\"{u}r Extraterrestrische  Physik (MPE), National Astronomical Observatory  of China,  New Mexico State  University, New York University, University of Notre Dame, Observat\'{o}rio Nacional / MCTI,  The  Ohio  State  University,  Pennsylvania  State  University, Shanghai Astronomical Observatory, United Kingdom Participation Group, Universidad  Nacional   Aut\'{o}noma  de  M\'{e}xico,   University  of Arizona,  University  of  Colorado   Boulder,  University  of  Oxford, University of Portsmouth, University  of Utah, University of Virginia, University   of  Washington,   University  of   Wisconsin,  Vanderbilt University, and Yale University.

\section{Data availability}
The data underlying this article will be shared on reasonable request to the corresponding author.

\bibliographystyle{mnras}
\bibliography{master}
\bsp
\label{lastpage}
\end{document}